\documentclass{ws-procs9x6}

\begin{document}

\title{The Scalar Sector and the $\eta \rightarrow 3\pi$ problem}

\author{A. ABDEL-REHIM$^a$, D. BLACK$^b$, A.~H. FARIBORZ$^c$ 
AND J. SCHECHTER$^d$\footnote{\uppercase{S}peaker}}

\address{(A) Physics Department, \\ 
Syracuse University, \\
Syracuse, NY 13244-1130,USA.\\
E-mail: abdou@phy.syr.edu}  

\address{(B) Theory Group, Jefferson Laboratory, \\
12000 Jefferson Avenue, \\
Newport News, Virginia, 23606, USA. \\
E-mail: dblack@jlab.org} 

\address{(C) Department of Mathematics/Science, \\
State University of New York Institute of Technology, \\
Utica, NY 13504-3050, USA. \\
E-mail: fariboa@sunyit.edu}

\address{(D)Physics Department, \\
Syracuse University, \\
Syracuse, NY 13244-1130, USA. \\
E-mail: schechte@phy.syr.edu}

\maketitle

\abstracts{
First, recent work on light scalar mesons, which is of possible
interest in connection with the strong coupling region of QCD,
is briefly discussed. Then a very short highlighting of a
paper concerned with an application to the $\eta\rightarrow 3\pi$
problem is presented.}

\section{Introduction}

 At very large
energy scales, the asymptotic freedom of QCD guarantees that a controlled
perturbation expansion is a practical tool.
 At very low energy scales, for example close to the
threshold of $\pi \pi$ scattering. the running QCD coupling constant is expected
to be large and perturbation theory is not expected to work. Fortunately, a
controlled expansion based on an effective theory
with the correct symmetry structure- Chiral Perturbation Theory\cite{GL}-
seems to work reasonably well. The new information about Strong Interactions which
this approach reveals is closely related to the spectrum and flavor "family"
properties of the lowest lying pseudoscalar meson multiplet and was, in fact,
essentially known before QCD.
 
\begin{figure}
\centering
\centerline{\epsfxsize=4.5in\epsfbox{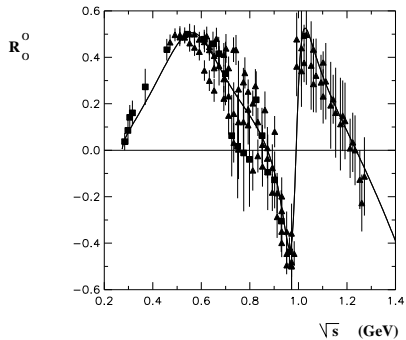}}
\caption{ Illustration of the real part of 
the pi pi scattering amplitude extracted from experimental data.}
\label{experiment}
\end{figure}

Clearly it is important to understand how far in energy above threshold 
the Chiral Perturbation Theory program will take us. To get a rough estimate
consider the experimental data for the real part of the $I=J=0$ $\pi \pi$
scattering amplitude, $R_0^0$ displayed in Fig.~\ref{experiment}. The chiral
perturbation series should essentially give a polynomial fit to this shape,
which up to about 1 GeV is crudely reminiscent of one cycle of a sine curve.

\begin{figure}
\centering
\vskip -2.0cm
\centerline{\epsfxsize=2.5in\epsfbox{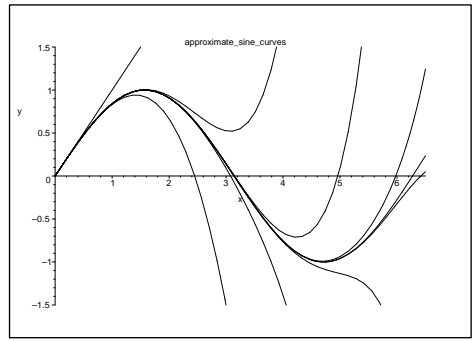}} 
\vskip -2.5cm
\caption{Polynomial approximations to one cycle of the sine curve.}
\label{polynomial}
\end{figure}

Now consider polynomial approximations to one cycle of the sine curve 
with various numbers of terms. These are illustrated in Fig.~\ref{polynomial}.
Note that each succesive term departs from the true sine curve right
after the preceding one. It is clear that something like eight terms
are required for a decent fit. This would correspond to seven loop order
of chiral perturbation theory and seems presently impractical.  

\section{Need for light scalar mesons}

Thus an alternative approach is indicated for going beyond threshold
of pi pi scattering up to about 1 GeV. The data itself suggests
the presence of s-wave resonances, the lowest of which is denoted the
"sigma". Physically, one then expects the practical range of
chiral perturbation theory to be up to about 450-500 MeV, just before the 
location of this lowest resonance.
In the last few years there have been studies
\cite{recentwork} by many authors which advance this picture. All
of them are "model dependent" but this is probably inevitable for
the strongly coupled regime of QCD.  For example \cite{HSS}, in a framework
where the amplitude is  computed from a non linear chiral Lagrangian
containing explicit scalars as well as vectors and pseudoscalars, the fit shown
in Fig.~\ref{experiment} emerges as a sum
of four pieces: i. the current algebra
"contact" term, ii. the $\rho$ exchange diagram iii. a non Breit Wigner
 $\sigma(560)$ pole
diagram and exchange, iv. an $f_0(980)$ pole in the background produced  
 by the other three. It is not just a simple sum of Born graphs but includes
the approximate unitarization features of the non Breit Wigner shape of the sigma 
and a Ramsauer Townsend mechanism which reverses the sign of the $f_0(980)$. 
Also note that i. and ii. provide very substantial background to the sigma
pole, partially explaining why the sigma does not "jump right out" of
various experimental studies. Qualitative agreement with this approach is obtained by
K-matrix unitarization of the two flavor linear sigma model \cite {ASII}
and three flavor linear sigma model \cite{BFMNS} amplitudes.

Workers on scalar mesons entertain the hope that, after the revelations
about the vacuum structure of QCD confirmed by the broken chiral symmetric
treatment of the pseudoscalars, an understanding of the next layer 
of the "strong interaction onion"
will be provided by studying the light scalars. An initial question is
whether the light scalars belong to a flavor $SU(3)$ multiplet as the
underlying quark structure might suggest. Apart from the $\sigma(560)$,
the $f_0(980)$ and the isovector $a_0(980)$ are fairly well established. This leaves
a gap concerning the four strange- so called kappa- states. This question is
more controversial than that of the sigma state . In the unitarized 
non linear chiral       
Lagrangian framework one must thus consider $pi-K$ scattering.
 In this case the low energy amplitude is taken\cite{BFSS}
to correspond to the sum of a current algebra contact diagram,
vector $\rho$ and $K^*$ exchange diagrams and scalar $\sigma(560)$,
$f_0(980)$ and $\kappa(900)$ exchange diagrams. The situation in the
interesting $I=1/2$ s-wave channel turns out to be very analogous to
the $I=0$ channel of s-wave $\pi\pi$ scattering. Now a non Breit Wigner 
$\kappa$ is required to restore unitarity; it
plays the role of the $\sigma(560)$ in the $\pi\pi$ case. It was found that
a satisfactory description of the 1-1.5 GeV s-wave region is also
obtained by including the well known $K_0^*(1430)$ scalar
resonance, which plays the role of the $f_0(980)$ in the $\pi\pi$
calculation. As in the case of the sigma, the light kappa seems hidden by
background and does not jump right out of the initial analysis of the 
experimental data.

 Thus the nine states
 associated with the $\sigma(560)$, $\kappa(900)$,
$f_0(980)$ and $a_0(980)$ seem to be required in order to fit experiment   
in this chiral framework. What would their masses and coupling constants suggest
about their quark substructure if they were assumed to comprise an SU(3) nonet
\cite{putative}?
 Clearly the mass ordering of the various
states is inverted compared to the "ideal mixing"\cite{Okubo}
scenario which approximately holds for most meson nonets. This means that
a quark structure for the putative scalar nonet of the form  $N^b_a \sim q_a{\bar
q}^b$ is unlikely since the mass ordering just corresponds to counting the
number of heavier strange quarks. Then the nearly degenerate $f_0(980)$ and $a_0(980)$
which must have the structure $N^1_1 \pm N^2_2$ would be lightest rather
than heaviest. However the inverted
ordering will agree with this counting if we assume that the scalar mesons   
are schematically constructed as $N^b_a \sim T_a{\bar T}^b$ where $T_a
\sim \epsilon_{acd}{\bar q}^c{\bar q}^d$ is a "dual" quark (or
anti diquark). This interpretation is strengthened by consideration
\cite{putative} of the scalars' coupling constants to two
pseudoscalars. Those couplings depend on the value of a mixing angle, $\theta_s$
between $N^3_3$ and $(N^1_1-N^2_2)/{\sqrt 2})$. Fitting 
 the coupling constants
to the treatments of $\pi\pi$ and $K\pi$ scattering gives a mixing angle such that
 $\sigma \sim N^3_3 +$ "small"; $\sigma$(560) is thus
a predominantly non-strange particle in this picture. Furthermore   
the states $N^1_1 \pm N^2_2$ now would each predominantly contain two extra
 strange quarks and would be expected to be heaviest.
  Four quark pictures of various types have been sugggested as arising from
spin-spin interactions in the MIT bag model\cite{Jaffe}, 
unitarized quark models\cite{uqm} and
 meson-meson interaction models\cite{mmim}.

There seems to be another interesting twist to the story of the light scalars. 
 The success of the phenomenological quark model suggests
that there exists, in addition, a nonet of ``conventional"
p-wave $q{\bar q}$ scalars in
the energy region above 1 GeV. The experimental candidates for these states
are $a_0(1450)(I=1)$, $K_0^*(1430)(I=1/2)$ and for $I=0$,
$f_0(1370)$, $f_0(1500)$ and $f_0(1710)$. These are enough for a full
nonet plus a glueball. However it is puzzling that the strange
$K_0^*(1430)$ isn't noticeably heavier than the non strange $a_0(1450)$
and that they are not lighter than the corresponding spin 2 states.
These and another puzzle may be solved in a natural way\cite{BFS}
if the heavier p-wave scalar nonet mixes with a lighter $qq{\bar q}
{\bar q}$ nonet of the type mentioned above. The mixing mechanism makes  
essential use of the "bare" lighter nonet having an inverted mass ordering
while the heavier "bare" nonet has the normal ordering.
A rather rich structure involving
the light scalars seems to be emerging. At lower energies one may consider
as a first approximation, "integrating out" the heavier nonet and retaining
just the lighter one.

\section{Effect of light scalars in $\eta\rightarrow 3\pi$}

\begin{figure}[htbp]   
\begin{center}
\epsfxsize = 5.0cm
\ \epsfbox{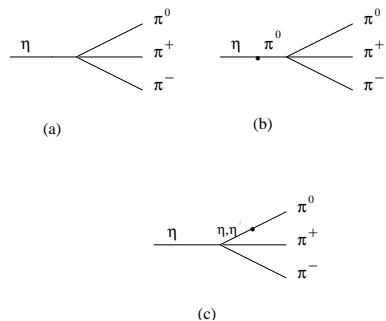}
\end{center}
\caption
{Feynman diagrams representing the pseudoscalar meson contribution
to the decay $\eta \rightarrow \pi^+ \pi^- \pi^0$ .}
\label{psdiagrams}
\end{figure}

\begin{figure}[htbp]
\begin{center}
\epsfxsize =5.0cm   
\ \epsfbox{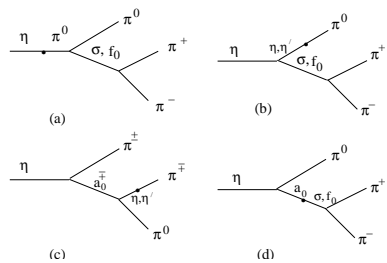}
\end{center}
\caption
{Feynman diagrams representing the scalar meson contributions
to the decay $\eta \rightarrow \pi^+ \pi^- \pi^0$ .}
\label{scdiagrams}
\end{figure}

Historically, this isospin violating process has been important as a
relatively clean test of the effective chiral Lagrangian approach and
as a source of information on the quark mass difference $m_d -m_u$.
As for the experimental status, the shape of the Dalitz plot
for the $\pi^+\pi^-\pi^0$ mode agrees with chiral models. The experimental
width for this mode \cite{H} is $267\pm 25$ eV. On the other hand,
the tree level chiral result (which might be expected to be accurate to within 
25 per cent or so) is $106$ eV while the one loop theoretical number
\cite{Gasser85} is $160\pm 50$ eV. A correction due to final state
interactions, but outside the chiral perturbation expansion, yields
\cite{Kambor96} $209\pm 20$ eV,

In ref.\cite{ABFS} we studied the effects of light scalars at tree level
for this still interesting reaction. The calculation used a chiral
Lagrangian of pseudoscalars, vectors and scalars, with the scalar masses
and coupling constants taken from ref.\cite{putative} mentioned above
and from\cite{FS}. We included non minimal derivative terms of
pseudoscalars but will only describe here the results for the minimal
pseudoscalar model with scalars and neglect of vectors. The 16 Feynman diagrams 
are shown in Fig. \ref{psdiagrams} and Fig. \ref{scdiagrams}. 
The "driving mechanism" for this decay is the non zero value of $m_d-m_u$
which results in isospin violating bilinear pseudoscalar terms
$\eta,\eta'-\pi^0$, bilinear scalar terms $f_0.f_0'-a_0^0$ as well
as the quadrilinear term in (a) of Fig.\ref{psdiagrams}.

The pseudoscalar diagrams in Fig. \ref{psdiagrams}
(with the minimal chiral Lagrangian) predict a width
of 106 eV for the $\pi^+\pi^-\pi^0$ mode. A priori
one might expect important contributions from
the sigma exchange diagrams in (a),(b) and (c) of Fig. \ref{scdiagrams},
since the $\sigma(560)$ propagators will not be too far from their mass
shells. However the signs are such that these three leading
scalar contributions almost cancel each other. The net result
is a total increase of the tree width by about 13 per cent to about 120 eV.
This is based on the fits we obtained for the coupling constants which appear.
 As discussed in section IV of \cite{ABFS} there may be some room
to increase this, owing to the delicate partial cancellations.  
The result is in the right direction but does illustrate
that some care is needed in treating this simple looking reaction.

 We are grateful to M. Harada, P. Herrera-Siklody and F.
Sannino for very helpful discussions. We would like to thank
Professors K. Yamawaki, Y. Kikukawa and M. Harada for all their hard work
in organizing this stimulating conference.
The work of A. A-R. and J.S. has been supported in part by the US DOE
under contract DE-FG-02-85ER 40231.  D.B. wishes to acknowledge
support from the Thomas Jefferson National Accelerator Facility
operated by the Southeastern Universities Research Association (SURA)
under DOE Contract No. DE-AC05-84ER40150.
The work of A.H.F. has been supported by grants from the State
of New York/UUP Professional Development Committee, and the 2002
Faculty Grant from the School of Arts and Sciences, SUNY Institute
of Technology.

\end{document}